\documentclass[aip,pof,reprint,amsmath,amssymb,floatfix]{revtex4-1}

\usepackage{graphicx}
\usepackage{dcolumn}
\usepackage{bm}

\providecommand{\gtrsim}{\:\raisebox{.25ex}{$>$}\hspace*{-.75em}
\raisebox{-.93ex}{$\sim$}\:}
\providecommand{\lesssim}{\:\raisebox{.25ex}{$<$}\hspace*{-.75em}
\raisebox{-.93ex}{$\sim$}\:}

\newlength{\myfigwidth}
\begin{document}


\title{Centered rarefaction wave with a liquid-gas phase transition in the approximation of ``phase-flip'' hydrodynamics} 


\author{Mikhail M. Basko}
\email{mmbasko@gmail.com}
\homepage{http://www.basko.net}
\affiliation{Keldysh Institute of Applied Mathematics, Miusskaya square 4, 125047 Moscow, Russia}

\date{\today}

\begin{abstract}
It is proposed to evaluate the effects of thermodynamic metastability on fluid dynamics by comparing two different ideal-hydrodynamics solutions --- one obtained with the fully equilibrium equation of state using the Maxwell construction, and the other in what we call the phase-flip approximation. The latter is based on the assumption of instantaneous decay of metastable states upon reaching the spinodal. The proposed method is applied to the classical problem of the centered rarefaction wave by expansion into vacuum, for which exact analytical solutions exist in both approximations. It is shown that the rapid decay of metastable states leads to the formation of a rarefaction shock in the expanding flow. Implications for the laser-heating experiments are discussed.
\end{abstract}


\maketitle 

\section{Introduction}

A typical situation in the majority of laser-matter experiments is where a laser beam heats up a surface layer of a solid (liquid) sample, which then begins to expand into vacuum or a low-pressure environment. In many cases, important for both practical applications and fundamental research, the expanding material passes through the liquid-gas phase coexistence region in the thermodynamic parameter space. As characteristic examples, two different groups of experiments with ultra-short laser pulses \cite{Linde_Sokow-Tinten.1997, Vinokhodov_Koshelev.2016} can be pointed out. In both these examples, an adequate description of fluid dynamics in the two-phase region proved to be crucial for the correct interpretation of the observed results \cite{Sokol-Tint_Bialkowski.1998, Basko_Krivokor.2017}.

Any attempt to employ the one-fluid hydrodynamics for modeling such experiments stumbles over a dilemma of what equation of state (EOS) ought to be used in the two-phase region below the critical point of the liquid-gas phase transition. One approach, which is fully self-consistent and poses neither conceptual nor technical difficulties, is to employ the \emph{fully equilibrium} (EQ) EOS, obtained by applying the well-known Maxwell rule. The EQ EOS was successfully used \cite{Sokol-Tint_Bialkowski.1998, Inogamov_Anisimov.1999} to explain the Newton rings observed in Ref.~\onlinecite{Linde_Sokow-Tinten.1997}, but only partially so by modeling \cite{Basko_Krivokor.2017} the fragmentation dynamics of liquid tin microspheres, hit by picosecond laser pulses \cite{Vinokhodov_Koshelev.2016}.

At the same time, an obvious flaw of the EQ hydrodynamics is the tacit assumption that the boiling of liquid (condensation of vapor) begins immediately upon crossing the outer boundary of the phase coexistence region --- the \emph{binodal}. In reality, especially in the laser-driven experiments where the hydrodynamic time scale can be as short as $\lesssim 0.1$~ns, the fluid elements can penetrate rather deeply into the metastability region and boil up (condense) quite close to the boundary of absolute thermodynamic instability --- the \emph{spinodal}. To remedy this deficiency of the EQ hydrodynamics, we propose an approach, based on comparing the EQ solution for a given problem with an alternative hydrodynamic solution for the same problem, obtained in what we call the \emph{phase-flip} (PF) approximation.

In the PF approximation we assume that each fluid element obeys the metastable (MS) EOS all the way down to the spinodal, where an instantaneous and irreversible relaxation to the EQ EOS takes place. The justification for this assumption is given in Section~\ref{s:pf}. On the one hand, the PF approximation is simpler than other models invoking the kinetics of phase transition (see, for example, Ref.~\onlinecite{Saurel_Petitpas.2008} and references therein) because one is allowed to stay within the framework of the one-fluid ideal hydrodynamics. On the other hand, allowing for maximum possible penetration into the region of metastable states, it can be expected to provide a good estimate for the maximum possible influence of metastability on the flow pattern under specific experimental conditions.

This paper is devoted to the application of the PF approach to one of the classical flows in the ideal hydrodynamics --- the centered rarefaction wave by expansion into vacuum \cite{Courant_Friedrichs1977, LL-H87, Zeldovich_Raizer2012}. Although the principal ideas behind the PF approximation have been discussed earlier \cite{Saurel_Petitpas.2008}, it is, to the best of our knowledge, for the first time that an exact analytical PF solution is presented in full detail (Section~\ref{s:crw}) and applied to draw specific qualitative and quantitative conclusions in regard to certain types of laser-driven experiments. This is done by making use of a specific form of the generalized van der Waals EOS introduced in Section~\ref{s:GW}, which is reasonably well suited for the description of real materials.

\section{Liquid-gas phase diagram for the generalized van der Waals equation of state \label{s:GW}}

All the numerical results in this paper are obtained with a generalized version of the van der Waals equation of state (GWEOS), where the power exponent $n$ in the attractive term is treated as a free parameter \cite{Martynyuk1991, Martynyuk1993, Basko2018}. This variant of a two-phase EOS preserves the convenient property of the original van der Waals EOS that it can be cast in the universal reduced form, where the dimensionless quantities $\rho$, $\theta$ and $p$ are, respectively, the density, temperature and pressure normalized by their values $\rho_{cr}$, $T_{cr}$ and $P_{cr}$ at the critical point; the latter can be considered as three independent free parameters of GWEOS taken, for example, from experiment.

In its reduced form,  GWEOS can be presented as
\begin{eqnarray}\label{GW:p(v,tet)=}
  p(v,\theta) &=& \frac{\alpha \theta}{v-\kappa^{-1}} -\frac{\kappa}{v^n},
  \\ \label{GW:e(v,tet)=}
  e(v,\theta) &=& c_V\alpha \theta -\frac{1}{2} \kappa(\kappa-1) v^{1-n},
  \\ \label{GW:s(v,tet)=}
  s(v,\theta) &=& \alpha \left[ c_V\left(1+ \ln\theta\right)
  +\ln\left(v-\kappa^{-1}\right) \right],
\end{eqnarray}
where $v = \rho^{-1}$ is the reduced specific volume, $e$ is the specific internal energy normalized by $P_{cr}/\rho_{cr}$, $s$ is the specific entropy normalized by $P_{cr}/\left(\rho_{cr}T_{cr}\right)$, and
\begin{equation}\label{GW:alpha=}
  c_V >0, \quad \alpha=\kappa-\kappa^{-1}, \quad \kappa =\frac{n+1}{n-1} > 1
\end{equation}
are dimensionless constants. The square of the reduced sound velocity, normalized by $P_{cr}/\rho_{cr}$, is given by
\begin{equation}\label{GW:cs2=}
  c_s^2 \equiv \left(\frac{\partial p}{\partial \rho}\right)_s =
  \frac{\left(1+c_V^{-1}\right) \alpha\, \theta v^2}{\left(v-\kappa^{-1} \right)^2} -\frac{n\kappa}{v^{n-1}}.
\end{equation}
Physically meaningful are the values $n>1$, implying $\kappa >1$. The original van der Waals EOS is recovered for $n=2$. Note that GWEOS is defined over a finite density interval $0 < \rho < \kappa$, i.e.\ for $v> \kappa^{-1}$.

Beside $\rho_{cr}$, $T_{cr}$ and $P_{cr}$, GWEOS has two more free parameters, namely, the heat capacity at constant volume $c_V$ and the power exponent $n$. Since $c_V$ is assumed to be constant, we use its ideal-gas value in the limit of $v\to \infty$; for the monoatomic substances $c_V=3/2$. A physically sensible recipe for evaluation of $n$ could be by fitting the
experimental values of the dimensionless ratio
\begin{equation}\label{GW:Lam=}
  \Lambda = \frac{E_{coh}}{T_{cr}} = \frac{(n+1)^{n+1}}{4n(n-1)^n},
\end{equation}
where $E_{coh}$ is the cohesive energy per atom (molecule). For a wide variety of monoatomic and diatomic substances the measured $\Lambda$ values \cite{Grigoriev1997} fall in the range $\Lambda \approx 4.0$--5.3, which implies $n = 1.4$--1.65. Somewhat different physical considerations \cite{Martynyuk_B1995} have led to practically the same values of $n$.  Accordingly, in this work we adopt a universal value of $n=1.5$. Note that this is one aspect where GWEOS is superior to the original van der Waals EOS, which yields a not quite realistic value of $\Lambda =3.375$.

\begin{figure}
\includegraphics[width=\myfigwidth]{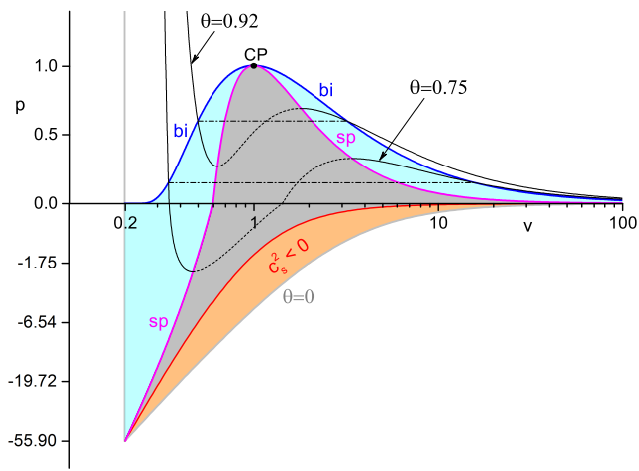}
\caption{(Color online) GWEOS with $n=c_V=1.5$ on the thermodynamic $(v,p)$ plane. Metastable states (shaded cyan) lie between the binodal $bi$ (blue) and the spinodal $sp$ (magenta) curves. The area between the spinodal $sp$ and the cold curve $\theta=0$ (shaded grey and grey-orange) is the region of absolute thermodynamic instability. The region of non-hyperbolicity with $c_s^2 <0$ is shaded grey-orange. Two isotherms $\theta=0.75$ and $\theta=0.92$ are shown as thin black curves in both the MS (solid) and EQ (dash-dotted) versions of GWEOS. Note that for $p<0$ the vertical axis has a different scale [linear in $\ln(1-p)$], which explains the kinks on isotherms and the spinodal by crossing the $p=0$ line. \label{f:1}}
\end{figure}

Main features of GWEOS for $n=1.5$ ($\kappa=5$) are illustrated in Fig.~\ref{f:1}. Below the critical point $CP$ at $p=v=\theta=1$ any isotherm (\ref{GW:p(v,tet)=}) with $\theta < 1$ has a segment $v_{sp,l}< v< v_{sp,g}$ of absolute thermodynamic instability, where $\partial p(v,\theta)/\partial v >0$. On the $(v,p)$ and $(v,\theta)$ planes the region of absolute instability lies below the spinodal curve defined by
\begin{eqnarray}\label{GW:p_sp(v)=} &&
  p=p_{sp}(v) = \frac{(n+1)v-n}{v^{n+1}}, \\ \label{GW:tet_sp(v)=} &&
  \theta = \theta_{sp}(v) = \frac{1}{v^{n+1}}
  \left(\frac{v-\kappa^{-1}}{1-\kappa^{-1}} \right)^2.
\end{eqnarray}
For any $p<1$, the liquid, $v= v_{sp,l}(p)< 1$, and the vapor, $v= v_{sp,g}(p)> 1$, branches of the spinodal are found as the two roots of Eq.~(\ref{GW:p_sp(v)=}). According to the basic thermodynamic principles, thermodynamic states, calculated from Eqs.~(\ref{GW:p(v,tet)=})--(\ref{GW:cs2=}) inside the interval $v_{sp,l}< v< v_{sp,g}$, cannot be realized in nature.

Equilibrium thermodynamics tells us \cite[\S 81,83]{LL-SP96} that, for any given $\theta< 1$, the liquid-vapor phase transition occurs at a fixed pressure $p<1$ across a finite interval $v_{bi,l}< v< v_{bi,g}$ of the phase coexistence. The functions $v= v_{bi,l}(p)< 1$ and $v= v_{bi,g}(p)> 1$, obtained by applying the Maxwell rule \cite[\S 84,85]{LL-SP96} to isotherms (\ref{GW:p(v,tet)=}), make up, respectively, the liquid and the vapor branches of the binodal curve $bi$ in Fig.~\ref{f:1}. Accordingly, the Maxwell construction applied to Eqs.~(\ref{GW:p(v,tet)=})--(\ref{GW:cs2=}) yields the EQ version of GWEOS for the entire phase coexistence region below the binodal. The full set of formulae for calculating the EQ GWEOS is given in Ref.~\onlinecite{Basko2018}. Note that one always has $v_{bi,l}(p)< v_{sp,l}(p)$ and $v_{bi,g}(p)> v_{sp,g}(p)$.

In the region between the binodal and spinodal curves (shaded cyan in Fig.~\ref{f:1}) the original Eqs.~(\ref{GW:p(v,tet)=})--(\ref{GW:cs2=}) are still physically meaningful: they represent the metastable thermodynamic states and provide the alternative MS option of GWEOS in this region. The region of non-hyperbolicity, where $c_s^2< 0$ and the hydrodynamic equations cease to be hyperbolic, lies always below the spinodal inside the region of absolute thermodynamic instability.

\begin{figure}
\includegraphics[width=\myfigwidth]{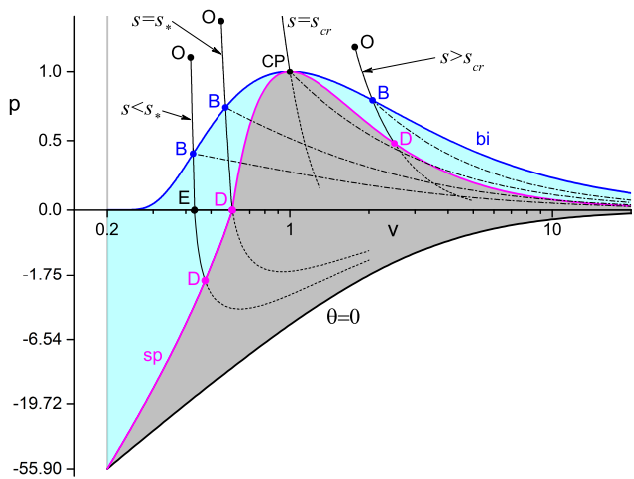}
\caption{(Color online) Four expansion isentropes on the $(v,p)$ plane for GWEOS with $n=c_V=1.5$. Upon crossing the binodal at point $B$, each isentrope splits into the metastable MS (black thin solid) and the equilibrium EQ (black dash-dotted) branches. Every MS branch finally reaches the spinodal at point $D$; its further continuation (black dots) into the instability region is assumed to be meaningless from the viewpoint of hydrodynamics. \label{f:2}}
\end{figure}

In this work we analyze a centered rarefaction wave into vacuum from a certain initial state $(v_0,p_0)$, denoted as point $O$ in Figs.~\ref{f:2}--\ref{f:4}. If the flow remains continuous, it evolves along the expansion isentrope starting at $O$. Figure~\ref{f:2} shows four examples of such isentropes for the GWEOS with $n= c_V=1.5$. When the origin point $O$ lies above the binodal, the isentrope crosses the binodal at a certain point $B$ with $v=v_B> v_0$. In ideal hydrodynamics, further expansion to $v> v_B$ can proceed either along the EQ branch of this isentrope (dash-dotted curves in Fig.~\ref{f:2}), or along its MS branch (thin solid curves) until the latter crosses the spinodal at a certain point $D$; we assume that below $D$ the MS branch of the expanding isentrope becomes physically meaningless. An important property of the EQ isentropes inside the phase coexistence region is that they never come out from under the binodal.

In GWEOS, the equation of an MS isentrope, passing through a certain point $(v_0,p_0)$ on the $(v,p)$ plane, is
\begin{equation}\label{GW:p_s(v)=}
  p = p_s(v) =\left(p_0 +\frac{\kappa}{v_0^n} \right) \left(\frac{v_0-\kappa^{-1}}{v-\kappa^{-1}} \right)^{\gamma} - \frac{\kappa}{v^n},
\end{equation}
where $\gamma = 1+c_V^{-1}$. If the condition $n< \gamma$ is fulfilled (as is the case for $n= c_V= 1.5$), any expanding MS isentrope earlier or later crosses the spinodal because the specific entropy $s$, calculated from Eq.~(\ref{GW:s(v,tet)=}) along the spinodal (\ref{GW:tet_sp(v)=}), monotonically increases with $v$. The latter provides one more argument in favor of GWEOS because the original van der Waals EOS with $c_V=1.5$ does not have this property.

Two of the four isentropes in Fig.~\ref{f:2} represent special cases. The critical isentrope $s=s_{cr}\equiv s(1,1)$ crosses both the binodal and the spinodal at the critical point $CP$; the metastable segment $BD$ is absent, and further expansion proceeds along the EQ-GWEOS branch. The special property of the isentrope $s= s_{\ast} \equiv s(v_{\ast},\theta_{\ast})$ is that it crosses the spinodal at the zero pressure $p_D =0$. The parameters of the crossing point (for $n=1.5$) are
\begin{equation}\label{GW:v_star=}
  v_{\ast} =\frac{n}{n+1} =0.6, \quad
  \theta_{\ast}= \frac{1}{4} \left(\frac{n+1}{n}\right)^{n+1} \approx 0.89652.
\end{equation}
This state corresponds to the highest possible temperature of a sharp liquid surface in vacuum. All isentropes with $s< s_{\ast}$ have an equilibrium metastable boundary with vacuum at point $E$, where $p_s(v_E) =0$ and $\kappa^{-1}< v_E <n/(n+1)$. For $s> s_{\ast}$ no stable or metastable boundary with vacuum at a non-zero density is possible.

\section{Phase-flip approximation \label{s:pf}}

In a hydrodynamic flow, the thermodynamic state of each fluid element traces a certain trajectory in the $(v,p)$ parametric plane. If such a trajectory crosses the binodal and enters the liquid-gas coexistence region, a non-trivial problem of the adequacy of the implemented EOS arises. The fully equilibrium EQ EOS, which is based on the Maxwell construction and would be the obvious choice, is poorly justified for fast dynamic processes because it assumes an infinitely fast relaxation to the full equilibrium immediately below the binodal, and, as a consequence, does not allow negative pressures. In practice, negative pressures do often play an important role, like in cavitation, or by providing a stabilizing mechanism for preservation of a sharp liquid-vacuum boundary up to relatively high surface temperatures --- an important issue for certain laser-driven experiments.

A better founded approach would be, once below the binodal, to proceed along the MS segment $BD$ of the thermodynamic trajectory and, simultaneously, keep track of the relaxation timescale $\tau_r$ from the MS to the EQ state. The MS-decay time $\tau_r$ must be compared with the dynamic timescale $\tau_h$ of the specific experiment. Clearly, so long as $\tau_r \gg \tau_h$, the MS branch should be a more adequate choice of EOS than the EQ one.

The dynamic timescale $\tau_h$ is determined either by the rate of hydrodynamic expansion or by such external factors as the boundary conditions, the rate of external heating, duration of observation, etc. By hydrodynamic expansion of laser-heated samples into vacuum the physically relevant timescale can be as short as $\tau_h \simeq \lambda/c_s \simeq 10^{-10}$~s, where $\lambda \simeq 1$~$\mu$m is the typical laser wavelength, and $c_s \simeq 10^6$~cm/s is the typical sound speed. For nanosecond laser pulses it can amount to $\tau_h \sim 10^{-9}$--$10^{-7}$~s. Under quasi-static conditions much larger values of $\tau_h$ are possible.

In superheated liquids the relaxation time $\tau_r$ can be evaluated on the basis of the theory of homogeneous bubble nucleation \cite{Skripov1974, Blander_Katz1975, Skripov_Skripov1979}. According to this theory, a superheated liquid state decays due to thermodynamic fluctuations, resulting in spontaneous creation of vapor bubbles with supercritical radii $r \geq r_c$ that subsequently grow in size. The critical radius $r_c$ in a given state $(p,T)$ is defined by the condition
\begin{equation}\label{pf:r_c=}
  p_{sat}(T) = p+ \frac{2\sigma}{r_c},
\end{equation}
where $p_{sat}(T)$ is the saturated vapor pressure, and $\sigma =\sigma(\rho,T)$ is the surface tension; here and throughout the text before Eq.~(\ref{pf:Hug=}) the thermodynamic variables are in conventional units. The energy barrier $W_c$ for the creation of a critical bubble is evaluated as
\begin{equation}\label{pf:W_c=}
  W_c = \frac{4\pi}{3} \sigma r_c^2 = \frac{16\pi\sigma^3}{3\left[p_{sat}(T)-p \right]^2}.
\end{equation}
Then the MS${}\rightarrow{}$EQ transition time can be evaluated as \cite{Faik_Basko.2012}
\begin{equation}\label{pf:tau_r=}
  \tau_r = \nu_V \left(V_c J \right)^{-1}, \quad J = B\, \exp\left( -\frac{W_c}{T}\right),
\end{equation}
where $0 \leq \nu_V= \nu_V(\rho,T) \leq 1$ is the volume fraction of vapor in the corresponding EQ state, $V_c =4\pi r_c^3/3$ is the volume of the critical bubble, $J$~[cm${}^{-3}$ s${}^{-1}$] is the spontaneous creation rate of bubbles with radius $r_c$, and $B = B(\rho,T)$ is a slowly varying preexponential factor.

The key factor, determining $\tau_r$, is the Gibbs exponent $\exp\left(W_c/T\right)$, which, in particular, yields $\tau_r =\infty$ at the binodal. As a thermodynamic trajectory penetrates the MS region and  approaches the spinodal, $\tau_r$ decreases and falls below $\tau_h$ somewhere close to the spinodal. Note that even in quasi-static laboratory experiments at $p \ll P_{cr}$ superheats up to $T \gtrsim 0.9 T_{cr}$ have been observed \cite{Skripov1974, Blander_Katz1975}.

Theoretical estimates for specific substances \cite{Martynyuk1977, Skripov_Skripov1979, Faik_Basko.2012} indicate that the transition from the condition $\tau_r \gg \tau_h$ to its opposite $\tau_r \ll \tau_h$ occurs near the spinodal in the course of a steep drop of $\tau_r$ by several orders of magnitude over a very narrow interval of temperatures $\Delta T/T_{cr} \simeq 1$--3\% (for quasi-isobaric trajectories with $p \ll P_{cr}$). In the immediate vicinity of the spinodal the model of homogeneous bubble nucleation breaks down \cite{Skripov1974}; at the spinodal itself the timescale  $\tau_r$ can roughly be estimated as not larger than a few tens of intermolecular collision times, i.e.\ $\tau_r \lesssim 10^{-12}$--$10^{-11}$~s in superheated liquids --- which is practically always shorter than the relevant hydrodynamic timescales $\tau_h$ in laser-heating experiments.

A very sharp transition from the situation with $\tau_r \gg \tau_h$ to that of $\tau_r \ll \tau_h$ suggests that one can make a simplifying assumption of $\tau_r = \infty$ along the initial stretch of the MS-EOS trajectory below the binodal until the transition point is reached, beyond which $\tau_r=0$ can be assumed. Earlier, a mathematical criterion for the moment of such transition was derived in Ref.~\onlinecite{Faik_Basko.2012} in the framework of the homogeneous nucleation theory. Because it is assumed to occur instantaneously, this MS${}\rightarrow{}$EQ transition may be called a \emph{phase flip (PF)}; in literature also the term \emph{phase explosion} is used \cite{Martynyuk1977}.

In this work we introduce a further simplification by assuming that the instantaneous and irreversible MS${}\rightarrow{}$EQ transition --- the phase flip --- occurs just at the intersection with the spinodal. More precisely, if the thermodynamic trajectory of a fluid element enters the MS region through the liquid binodal branch, the phase flip occurs when crossing the liquid branch of the spinodal; respectively, if the MS region is entered from the vapor side, it occurs upon reaching the vapor spinodal branch. We refer to this variant of hydrodynamic treatment of fluids with a liquid-gas phase transition as the \emph{PF hydrodynamics}.

\begin{figure}
\includegraphics[width=\myfigwidth]{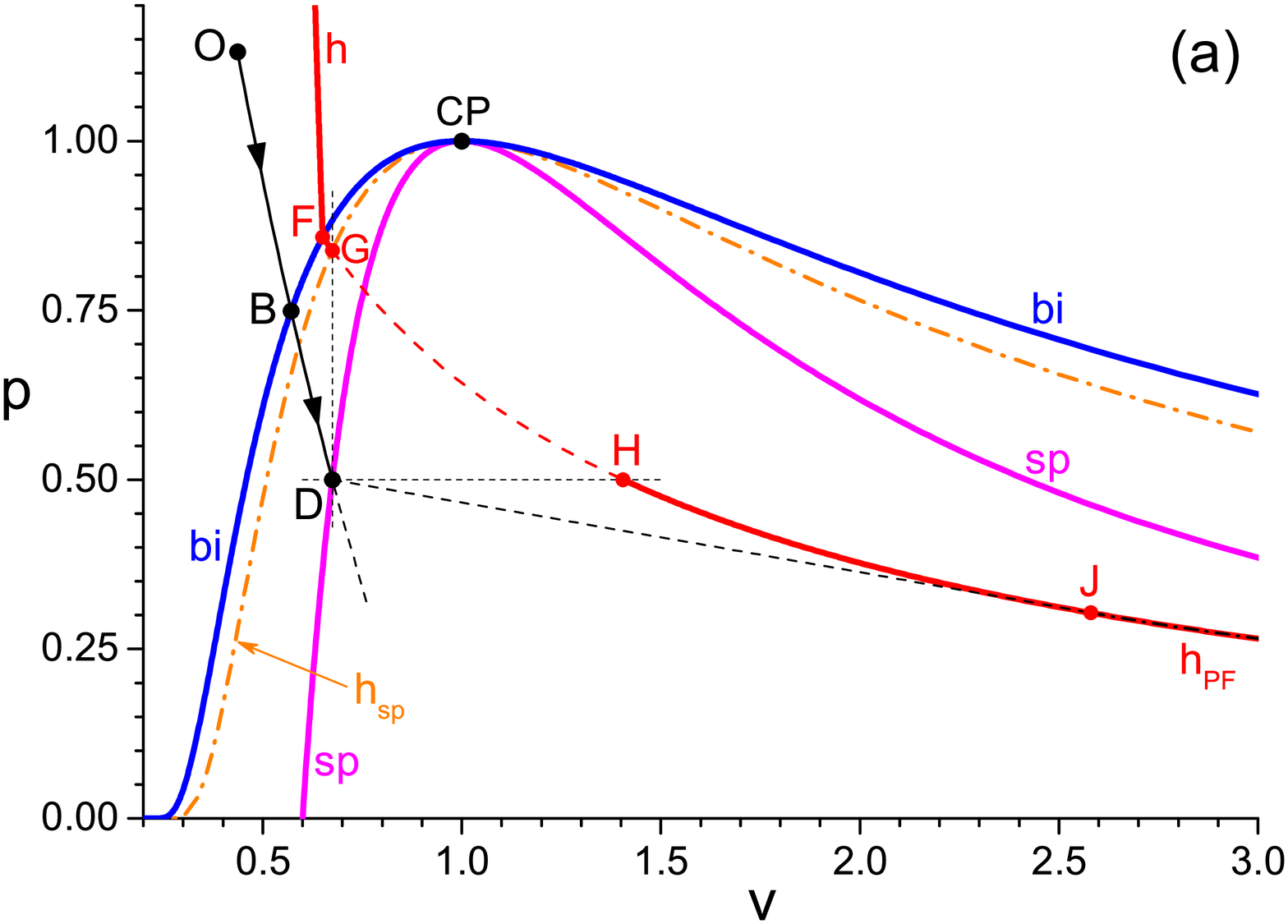}
\includegraphics[width=\myfigwidth]{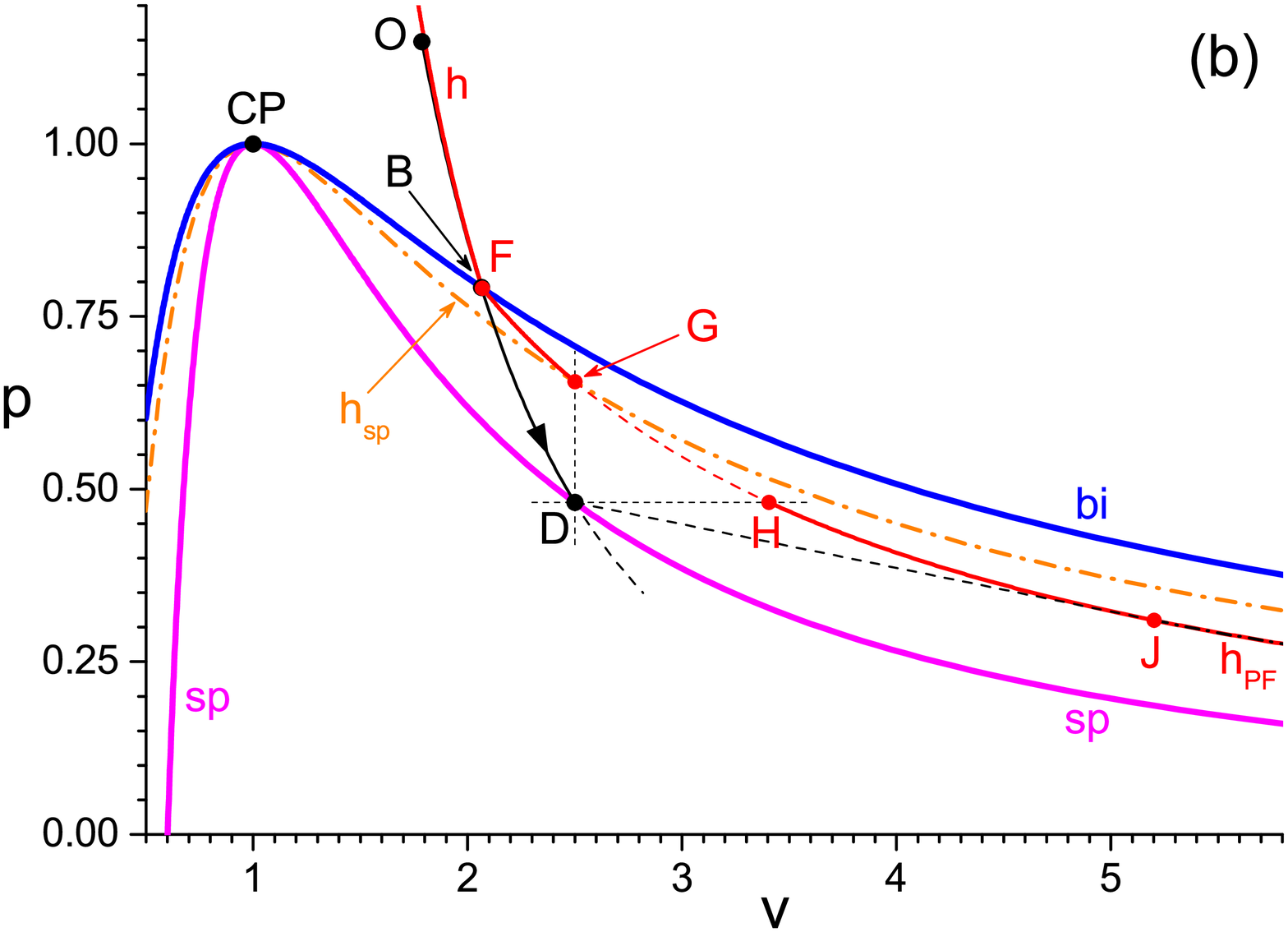}
\caption{(Color online) Two examples of the phase-flip Hugoniot  $hFGHJh_{PF}$ (red), calculated from Eq.~(\ref{pf:Hug=}) for the initial state  $D$ lying on the liquid (a) and the vapor (b) branches of the spinodal in GWEOS with $n=c_V=1.5$. In both cases the pre-flip state $D$ is presumed to be reached along an expansion isentrope $OBD$. On the upper graph (a) the isentrope $OBD$ is artificially tilted counterclockwise to avoid visual coalescence with the Hugoniot segment $h$ [as on graph~(b)]. Dash-dotted curve $h_{sp}$ (orange) is the locus of isochoric post-flip states $G$ as the initial state $D$ slides along the spinodal $sp$ (magenta). \label{f:3}}
\end{figure}

Generally, the dynamic response of a fluid to a jump in its thermodynamic properties would lead to a flow discontinuity, i.e.\ generally a phase flip should occur inside a PF shock front. In principle, it may be either a compression or a rarefaction PF shock. Conservation of mass, momentum and energy implies that, as in the theory of conventional shocks, the fluid states in front (initial) and behind (final) a PF shock must obey the Hugoniot relation \cite{Courant_Friedrichs1977, LL-H87}
\begin{equation}\label{pf:Hug=}
  e_2 -e_1 = \frac{1}{2} (p_1+p_2) (v_1-v_2).
\end{equation}
The initial state (point $D$ in Fig.~\ref{f:3}) is fully defined by its specific volume $v_1$: because it lies on the spinodal, we have $\theta_1= \theta_{sp}(v_1)$, while $p_1 = p(v_1,\theta_1)$, $e_1= e(v_1,\theta_1)$ and $s_1= s(v_1,\theta_1)$ are given by Eqs.~(\ref{GW:p(v,tet)=})--(\ref{GW:s(v,tet)=}). The final state $p_2 = p_{EQ}(v_2,\theta_2)$, $e_2 =e_{EQ}(v_2,\theta_2)$ and $s_2 =s_{EQ}(v_2,\theta_2)$ must be calculated from the EQ EOS  by invoking the Maxwell construction. Once the initial state $v_1$ is fixed and $v_2$ is varied, Eq.~(\ref{pf:Hug=}) defines a unique PF Hugoniot curve $p_2(v_2)$. Two characteristic examples of the PF Hugoniot with initial states on the liquid and the vapor spinodal branches are shown in Figs.~\ref{f:3}a and \ref{f:3}b.

Because the final state is calculated from a different EOS branch, the PF Hugoniot bears a close resemblance to the well-known detonation-deflagration Hugoniot curve \cite{Courant_Friedrichs1977, LL-H87}. Similar to the latter, the segment $GH$ between the vertical and horizontal lines, passing through the initial state $D$, has no physical meaning because the square of the mass flux \begin{equation}\label{pf:j^2=}
  j^2 =(\rho_1\eta_1)^2 =(\rho_2\eta_2)^2 =\frac{p_2-p_1}{v_1-v_2}
\end{equation}
across a PF front would be negative there; here $\eta_1$ and $\eta_2$ are, respectively, the flow velocities before and after the PF front in the reference frame comoving with this front. For the problem of adiabatic unloading into vacuum, studied in this work, only the segment below the $H$ point is relevant.

The phase flip at a constant density from $D$ to $G$ can be singled out as a special case: it occurs, for example, by a quasistatic volumetric heating (cooling) of an extended uniform liquid (vapor) volume at constant pressure \cite{Faik_Basko.2012}. In this case, when $v_1=v_2$, Eq.~(\ref{pf:Hug=}) yields a ``PF image'' $p_2(v_1)$ of the spinodal --- the curve $h_{sp}$ in Fig.~\ref{f:3}, which is the locus of final states $G$ when the initial state $D$ slides along the spinodal. Note that the $h_{sp}$ curve lies everywhere between the spinodal and the binodal [in both the $(v,p)$ and $(v,\theta)$ parametric planes], which implies that the isochoric explosive boiling of superheated liquid, as well as the rapid condensation of supersaturated vapor, are both accompanied by an upward jump in pressure and temperature.

In summary, the EQ and PF approximations represent two opposite extremes in the hydrodynamic description of fluid dynamics in the region of liquid-gas phase transition. Once both the EQ and the PF solutions are found for one and the same problem, a comparison between the two can provide valuable information about a possible impact of metastable states. Because the EQ option ignores metastable states altogether, whereas the PF approximation allows maximum penetration into the MS region, one can expect that such a comparison should provide a good measure for maximum possible dynamical effects of metastability.

\section{Centered rarefaction wave by expansion into vacuum  \label{s:crw}}

\subsection{Self-similarity ansatz}

The unloading of an initially uniform half-space bordering on vacuum is described by the one-dimensional self-similar solution for a centered rarefaction wave (CRW), where one family of characteristics in the $(x,t)$ plane is a set of straight lines originating from a single center \cite{Courant_Friedrichs1977, LL-H87, Zeldovich_Raizer2012}. A remarkable property of CRW is that it remains self-similar for any equation of state. If the rarefaction flow is continuous, it is also isentropic. If the initial condition is $\rho(t,x) =\rho_0$, $p(t,x) =p_0$, $\theta(t,x) =\theta_0$ for $x<0$ at $t=0$, than all the flow parameters are functions of a single independent variable $\xi=x/t$, and the solution takes the form
\begin{eqnarray}\label{crw:xi=u-c}
  \xi &\equiv&  \frac{x}{t} = u(\rho)-c_s(\rho), \\ \label{crw:u=}
  u(\rho) &=& \int\limits_{\rho}^{\rho_0} \frac{dp(\rho')}{\rho' c_s(\rho')} = \int\limits_{\rho}^{\rho_0} c_s(\rho')\,d\ln\rho',
\end{eqnarray}
where $u$ is the flow velocity, and $c_s$ is the isentropic sound velocity. In Eqs.~(\ref{crw:xi=u-c}) and (\ref{crw:u=}) both $p(\rho)$ and $c_s(\rho)$ are considered to be known functions of $\rho$ along the expansion isentrope
$s=s_0$. Once the integral in Eq.~(\ref{crw:u=}) is calculated for the given EOS, Eq.~(\ref{crw:xi=u-c}) defines in implicit form the functions $\rho(t,x)$ and $u(t,x)$.

For subsequent analysis it will be convenient to introduce the notion of the \emph{phase} of CRW as an arbitrary point on its profile with some fixed values of the material velocity $u$, density $\rho \equiv v^{-1}$, and other thermodynamic variables. Equation (\ref{crw:xi=u-c}) tells us that the phase is uniquely defined by the value of $\xi$, which, in its turn, can be called the \emph{phase velocity} because points with fixed $\rho$ and $p$  propagate in space with the velocity $\xi$. The head of the rarefaction wave, where $\rho= \rho_0$ and $u=0$, propagates into the unperturbed matter with the phase velocity
\begin{equation}\label{crw:xi_0=}
  \xi_0 =-c_s(\rho_0) \equiv -c_{s0}.
\end{equation}
The difference $\eta \equiv u-\xi$ is the material velocity with respect to a fixed wave phase, i.e.\ the \emph{phase-relative} flow velocity. One can prove that the absolute magnitude of the phase-relative velocity in any continuous adiabatic flow is equal to the local speed of sound $c_s$, as is expressed by Eq.~(\ref{crw:xi=u-c}) for the CRW case.

\subsection{The EQ solution}

\begin{figure}
\includegraphics[width=0.5\myfigwidth]{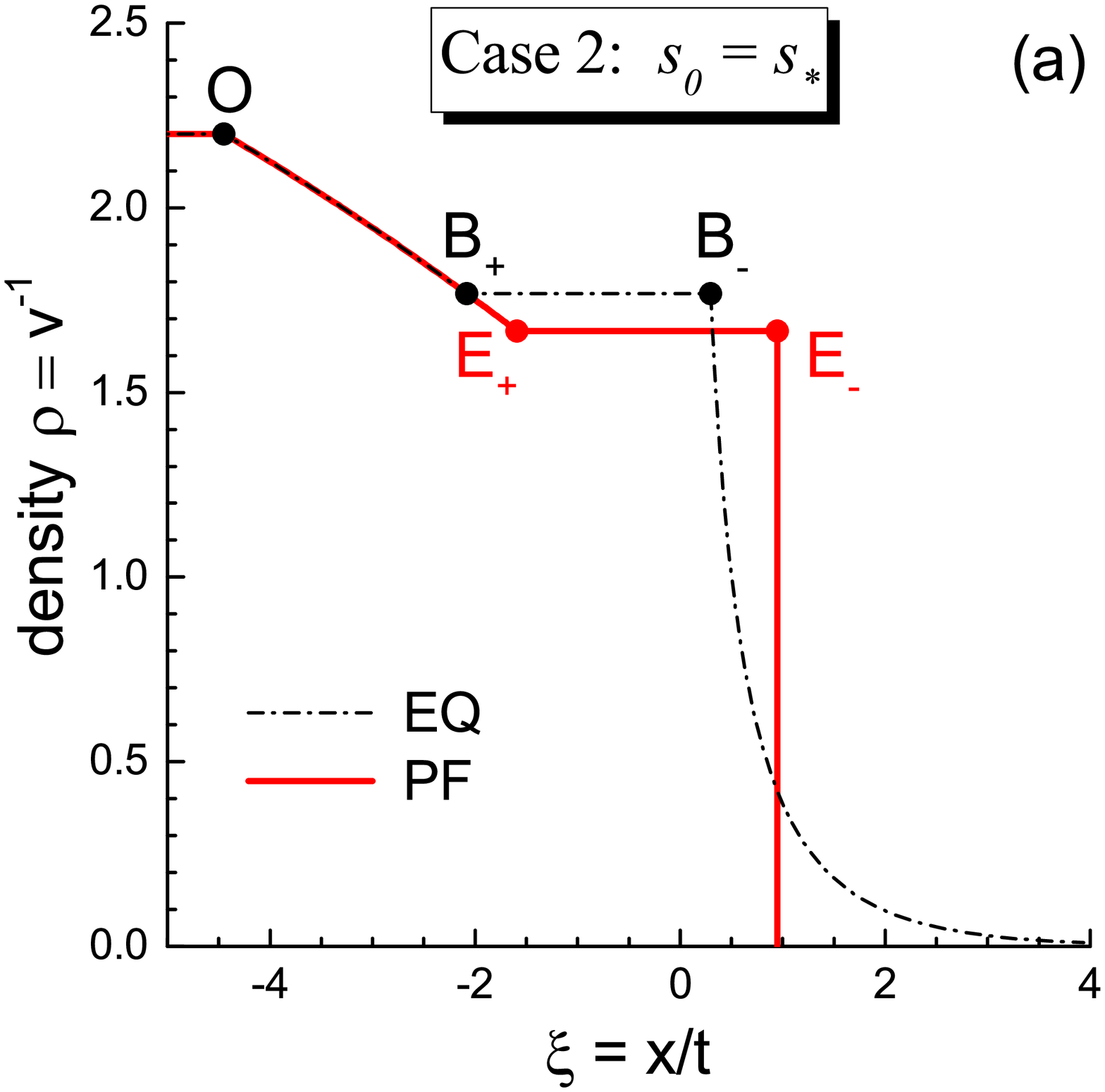}
\includegraphics[width=0.5\myfigwidth]{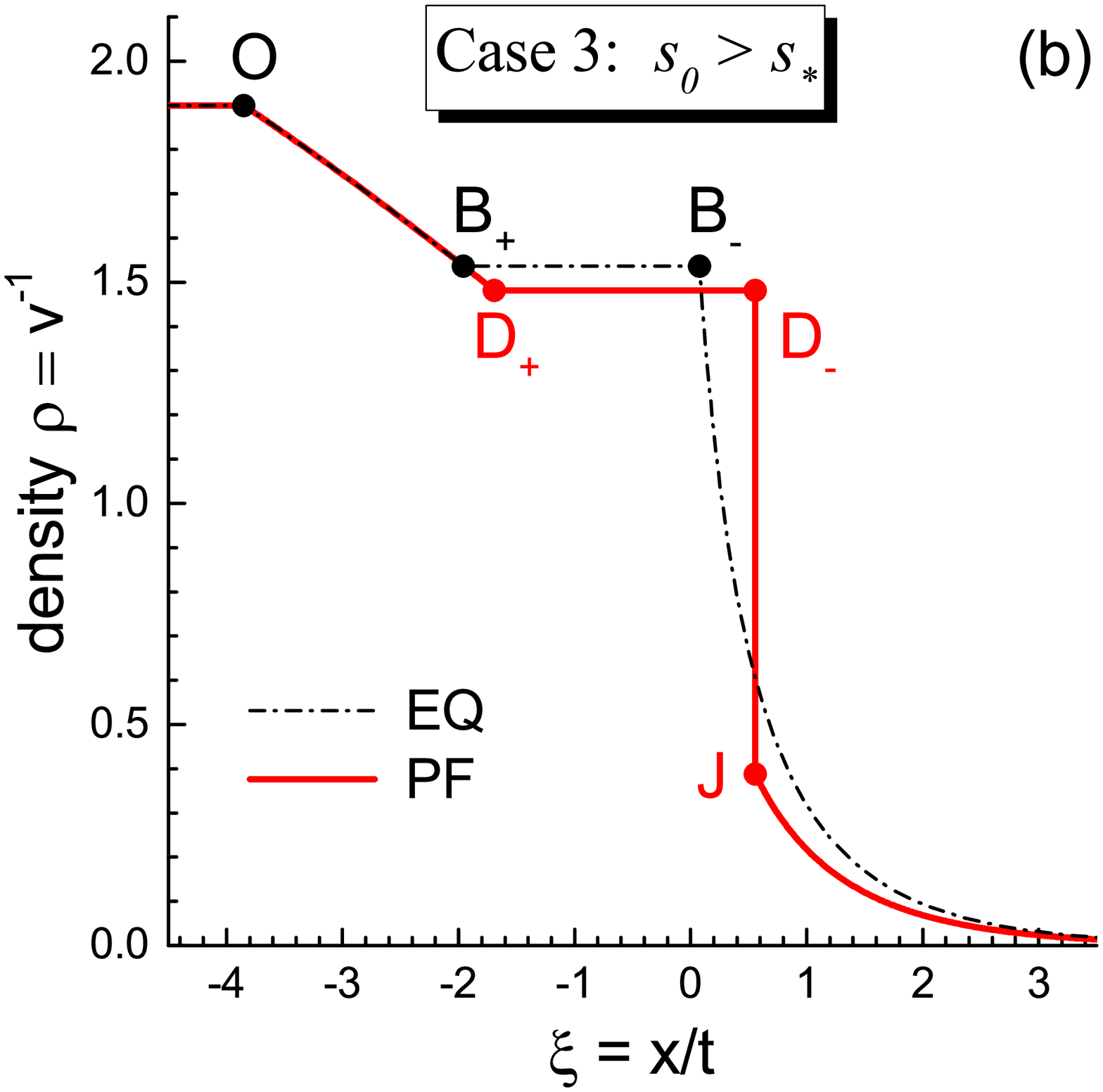}
\caption{(Color online) EQ (dash-dotted black) and PF (solid red) density profiles in the self-similar centered rarefaction wave, calculated for cases 2 (a) and 3 (b) in Table~\ref{t:1}. \label{f:4}}
\end{figure}

The CRW solution for the EQ EOS has been described in detail in Ref.~\onlinecite{Inogamov_Anisimov.1999}. It is continuous in space, and the thermodynamic trajectories of all fluid elements follow one and the same expansion EQ isentrope $s=s_0 \equiv s_{EQ}(v_0,\theta_0)$ defined by the initial state $O$, as is shown in Fig.~\ref{f:2}. Qualitatively, the spatial CRW profiles look similar for all unloading EQ isentropes. As an illustration, Fig.~\ref{f:4} shows the EQ density profiles $\rho(\xi)$ (black dash-dotted) for two different $s_0$ values. More data on the EQ-CRW solutions for five different values of $s_0$ are given in Table~\ref{t:1}.

\begin{table}
\caption{\label{t:1} Parameters of the EQ-CRW solutions for five representative cases with different values of the initial entropy $s_0$, calculated for GWEOS with $n=c_V=1.5$. For notation see the text.}
\begin{ruledtabular}
\begin{tabular}{lccccc}
 Case & 1  & 2 & 3 & 4 & 5  \\
  & $s_0< s_{\ast}$  & $s_0= s_{\ast}$ & $s_{\ast}\! <\! s_0\! <\! s_{cr}$ &
    $s_0= s_{cr}$ & $s_0> s_{cr}$  \\ [4pt]
\hline \\ [-4pt]
 $v_B$ & 0.27468 & 0.56574 & 0.65070  & 1 & 2.06392  \\
 $p_B$ & 0.01055 & 0.73952 & 0.85803 & 1 & 0.79175  \\
 $\theta_B$ & 0.54055 & 0.95166 & 0.97499 & 1 & 0.96226  \\
 $c_{sB_+}$ & 6.64741 & 2.87141 & 2.63834 & 2.23607 & 2.05383  \\
 $c_{sB_-}$ & 0.00925 & 0.49076 & 0.60481 & 0.91158 & 1.27836  \\
 $\Gamma_{B_-}$ & 0.03677 & 0.6091 & 0.6809 & 0.6909 & 0.9617  \\ [3pt]
 $v_D$ & 0.321 & 0.6 & 0.67482 & 1 & 2.5  \\
 $p_D$ & -11.948 & 0 & 0.5 & 1 & 0.48067  \\
 $\theta_D$ & 0.39186 & 0.89652 & 0.94169 & 1 & 0.83642  \\
 $c_{sD_+}$ & 2.97070 & 2.54066 & 2.46711 & 2.23607 & 1.77828  \\ [3pt]
 $\Delta\xi_{sh,EQ}$ & 6.63816 & 2.38065 & 2.03353 & 1.32448 & 0.77547  \\
 $\Delta\xi_{lvt,EQ}$ & 0.00034 & 0.2989 & 0.4118 & 0.6298 & 1.2294 \\
\end{tabular}
\end{ruledtabular}
\end{table}

A salient feature of the EQ-CRW profiles is the \emph{binodal shelf} --- a uniform layer $B_+ B_-$ with constant values of $u$, $\rho$ and other thermodynamic variables. The binodal shelf is formed at point $B$ (see Fig.~\ref{f:2}), where the unloading isentrope crosses the binodal at $\rho= \rho_B \equiv v_B^{-1}$ and the isentropic speed of sound $c_s$ experiences a jump from a higher upstream value $c_{sB_+}$, calculated along the MS-EOS branch, to a lower downstream value $c_{sB_-}$, calculated along the EQ-EOS branch. The flow velocity $u_B=u(\rho_B)$ on the shelf is given by Eq.~(\ref{crw:u=}).

Qualitatively, the formation of a binodal shelf can be explained as follows: while material enters the state $\rho=\rho_B$ from the upstream side with a phase-relative velocity $\eta_{B_+}= c_{sB_+}$, it leaves this state with a lower phase-relative velocity $\eta_{B_-}= c_{sB_-}$; hence, material accumulates in state $B$ at a rate $\rho_B(c_{sB_+}- c_{sB_-})> 0$, and the width of the resulting uniform layer grows in time at a speed
\begin{equation}\label{crw:Dxi_B=}
  \Delta \xi_{sh,EQ} \equiv \xi_{B_-}-\xi_{B_+} = c_{sB_+}- c_{sB_-} >0.
\end{equation}
That is to say, a single point $B$ on the $(v,p)$ plane is represented by a finite interval $\xi_{B_-}\leq \xi\leq \xi_{B_+}$ of the phase velocities.

At the boundary with vacuum, the phase velocity $\xi_e =u_e$ coincides with the flow velocity $u_e=u(0)$ calculated from Eq.~(\ref{crw:u=}). Normally, the integral in (\ref{crw:u=}) converges in the limit of $\rho \to 0$, though relatively slowly.  The layer $\xi_{B_-} <\xi< \xi_e$ downstream from the binodal shelf is composed of a disperse mixture of liquid and vapor. At low entropies $s_0 \lesssim s_{\ast}$, this \emph{liquid-vapor tail} exhibits a steep density drop, clearly manifested by Fig.~\ref{f:4}a. From Eqs.~(\ref{crw:xi=u-c}) and (\ref{crw:u=}) one calculates the effective width of this tail to be
\begin{equation}\label{crw:Dxi_lvt=}
  \Delta \xi_{lvt,EQ} \equiv\left| \frac{d\ln\rho}{d\xi}
  \right|^{-1}_{\xi =\xi_{B_-}+0} = c_{sB_-}\Gamma_{B_-},
\end{equation}
where $\Gamma_{B_-}$ is the value of the fundamental gasdynamic derivative \cite{Thomson1971}
\begin{equation}\label{crw:Gam=}
  \Gamma =1+\left(\frac{\partial \ln c_s}{\partial\ln\rho}\right)_s,
\end{equation}
calculated along the EQ branch of the expansion isentrope immediately below point $B$. In the present EOS, the tail width $\Delta \xi_{lvt,EQ}$ is a monotonically increasing function of the specific volume $v_B$ (see Table~\ref{t:1}) over a broad interval $0.2< v_B \lesssim 40$. As $p_B\to 0$ ($v_B \to 0.2$) on the liquid branch, $\Delta \xi_{lvt,EQ}$ very rapidly approaches zero, and the tail becomes physically insignificant --- which is expected to be true for any realistic EOS.

\subsection{The PF solution}

Depending on how the initial entropy $s_0$ compares with $s_{\ast}$, the PF-CRW solution can be of one of the two types, illustrated by Figs.~\ref{f:4}a and \ref{f:4}b respectively. When $s_0 \leq s_{\ast}$, the boundary condition $p(\rho_E) =0$ is satisfied at point $E$ (see Fig.~\ref{f:2}) with a non-zero density $\rho_E \equiv v_E^{-1} \geq 1+n^{-1}$ before the MS isentrope reaches the spinodal. In this case, displayed by Fig.~\ref{f:4}a, no phase flip occurs, the liquid-vapor tail is absent, and the boundary with vacuum remains a sharp surface of superheated liquid. Because material enters the boundary state $E$ from the upstream side with the phase-relative velocity $\eta_{E_+} =c_{sE}$, it accumulates there with the rate $\rho_E c_{sE}$ and forms a shelf $E_+E_-$ of width
\begin{equation}\label{crw:Dxi_E=}
  \Delta\xi_{sh,PF} \equiv \xi_{E_-}-\xi_{E_+} =c_{sE}.
\end{equation}
That is, for $-c_{s0}\leq \xi \leq \xi_{E_+}$ the PF solution is given by Eqs.~(\ref{crw:xi=u-c}) and (\ref{crw:u=}), while at $\xi_{E_+}< \xi< \xi_{E_-}$ the flow is uniform with $\rho=\rho_E$ and $u=u_e=u(\rho_E)$; the phase velocities at the two ends of the $E_+E_-$ shelf are $\xi_{E_+}= u_e-c_{sE}$ and $\xi_{E_-}\equiv \xi_e= u_e$. Parameters of state $E$ for the illustrative cases 1 and 2 (in case~2 points $E$ and $D$ coincide) are listed in Table~\ref{t:2}.

\begin{table}
\caption{\label{t:2} Parameters of the PF-CRW solutions for four representative cases from Table~\ref{t:1}. For notation see the text.}
\begin{ruledtabular}
\begin{tabular}{lcccc}
 Case & 1  & 2 & 3 & 5  \\
  & $s_0< s_{\ast}$  & $s_0= s_{\ast}$ & $s_{\ast} < s_0 < s_{cr}$ &
     $s_0> s_{cr}$  \\ [4pt]
\hline \\ [-4pt]
 $v_E$ & 0.27470 & 0.6 & --  & --  \\
 $\theta_E$ & 0.54046 & 0.89652 & -- & --   \\
 $c_{sE}$ & 6.64519 & 2.54066 & -- & --   \\ [3pt]
 $v_J$ & -- & -- & 2.57976 & 5.20258   \\
 $p_J$ & -- & -- & 0.30343 & 0.30995   \\
 $\theta_J$ & -- & -- & 0.82902 & 0.83164   \\
 $c_{sJ}$ & -- & -- & 0.82871 & 1.30758  \\
 $\Gamma_{J}$ & -- & -- & 0.8711 & 1.0074  \\
 $s_J-s_0$ & -- & -- & 0.17477 & 0.18982  \\
 $\nu_{m,J}$ & -- & -- & 0.31055 & 0.70077 \\ [3pt]
 $\Delta\xi_{sh,PF}$ & 6.64519 & 2.54066 & 2.25034 & 1.14995   \\
 $\Delta\xi_{lvt,PF}$ & 0 & 0 & 0.7219 & 1.3173   \\
 $\delta u_{sh}$ & 0.000436 & 0.15895 & 0.09289 & 0.36660 \\
\end{tabular}
\end{ruledtabular}
\end{table}

At higher initial entropies $s_0> s_{\ast}$, the unloading isentrope crosses the spinodal at point $D$ with $p_D>0$, and the vacuum boundary condition can only be satisfied after relaxation to the EQ EOS. Because the assumed instantaneous MS${}\rightarrow{}$EQ transition incurs a jump in pressure and/or density, it must take place inside a flow discontinuity, i.e.\ inside a PF shock front with the pre-shock state $(v_1,p_1) = (v_D,p_D)$ at point $D$, and the post-shock state $(v_2,p_2)$ on the corresponding PF Hugoniot displayed in Fig.~\ref{f:3}. In our case it will be a rarefaction shock with $p_2< p_D$ because the CRW evolves from a step-like initial pressure drop where all fluid elements are accelerated in one direction towards vacuum, i.e.\ the post-shock state $(v_2,p_2)$ must lie on the ``deflagration'' branch $HJh_{PF}$ of the PF Hugoniot.

Further on, we argue that the post-shock state must be exactly at the Chapman-Jouguet point $J$, where the ray $DJ$ is tangent to the PF Hugoniot, the outflow velocity $\eta_2$ from the shock front equals the local speed of sound $c_{sJ}$, and the mass flux
across the front
\begin{equation}\label{crw:j_J=}
  j\equiv \rho_2\eta_2 =j_{DJ} = \left( \frac{v_J-v_D}{p_D-p_J} \right)^{1/2} =\rho_J c_{sJ}
\end{equation}
is maximum for the given initial state $D$ \cite{Courant_Friedrichs1977, LL-H87}. Indeed, the conservation of mass and momentum dictates that, for any given $j$, the transition from the MS state $D$ to the EQ state 2 inside the shock front must proceed along the Rayleigh line
\begin{equation}\label{crw:p_tran(v)=}
  p(v) =p_D-(v-v_D)/j^2.
\end{equation}
If $j< j_{DJ}$, the line (\ref{crw:p_tran(v)=}) crosses the PF Hugoniot at two points, say at $J'$ with $v_{J'}<v _J$ and at $J''$ with $v_{J''}>v _J$. Physically meaningful would only be the first solution $J'$ because the kinetics of the irreversible MS${}\rightarrow{}$EQ transition would ensure that the final EQ state be already reached there. On the other hand, because the shock-outflow velocity $\eta_2$ would in this case be subsonic, $\eta_2< c_{sJ'}$, the zero pressure at the outer boundary would inevitably ``suck out'' the fluid from the shock front at $J'$ until the post-shock state $J$ with exactly the sonic outflow were reached; more detailed arguments to this point can be found in \S 20 of chapter~XI in Ref.~\onlinecite{Zeldovich_Raizer2012}. Parameters of the Chapman-Jouguet point $J$ for our illustrative cases~3 and 5 are listed in Table~\ref{t:2}. Case~4 with $s_0=s_{cr}$ is a degenerate one in the sense that points $B$, $D$ and $J$ coalesce with the critical point $CP$, and the PF and EQ solutions coincide. Note that the post-shock entropy $s_2(v_2)$ on the deflagration branch of the PF Hugoniot has its local maximum $s_J= s_2(v_J)$ at point $J$ \cite{Courant_Friedrichs1977}.

Having uniquely determined the post-shock state $J$, we notice that matter enters state $D$ at the rate $\rho_D c_{sD_+}$, and leaves it through the shock front at the rate $j_{DJ}= \rho_J c_{sJ}$. Normally (see Fig.~\ref{f:3}) the Rayleigh line $DJ$ slopes down less steeply than the MS isentrope at point $D$, which means that $\rho_D c_{sD_+}> j_{DJ}$ and material accumulates in state $D$ upstream from the PF shock. In result, the \emph{spinodal shelf} $D_+D_-$ of a uniform flow with $\rho=\rho_D$ is formed, whose width grows in time at a speed
\begin{equation}\label{crw:Delxi_sh,PF=}
  \Delta\xi_{sh,PF} \equiv \xi_{D_-}- \xi_{D_+} = c_{sD_+} -\frac{v_D}{v_J} c_{sJ}.
\end{equation}

The overall density profile of the PF solution for $s_0> s_{\ast}$ is displayed in Fig.~\ref{f:4}b. Similar to the EQ solution, the head of the rarefaction wave at $-c_{s0} \leq \xi \leq \xi_{D_+}$ is described by Eqs.~(\ref{crw:xi=u-c}) and (\ref{crw:u=}), where the sound speed $c_s(\rho)$ is calculated along the initial MS isentrope $s=s_0$. The flow velocity on the spinodal shelf at $\xi_{D_+}< \xi < \xi_{D_-}$ is $u_D=u(\rho_D)$. At $\xi =\xi_{D_-} =\xi_J$ the fluid passes through a rarefaction jump from state $D$ to state $J$, in which the entropy increases from $s_0$ to $s_J$, the density drops from $\rho_D=v_D^{-1}$ to $\rho_J=v_J^{-1}$, and the flow velocity jumps from $u_D$ to $u_D +c_{sJ}(1-v_D/v_J)$. When $s_{\ast}< s_0< s_{cr}$, the pure liquid, entering the PF shock, undergoes instantaneous partial vaporization (explosive boiling) inside the shock front; for $s_0> s_{cr}$ the pure (dry) vapor, entering the PF shock, undergoes instantaneous partial condensation. The values $\nu_{m,J}$ of the mass fraction of vapor in state $J$ behind the shock are listed in Table~\ref{t:2}.

Downstream from the PF shock we have a liquid-vapor tail at $\xi_J< \xi <\xi_e=u_e$, analogous to that in the EQ solution. There the flow is again continuous and isentropic, and the density and velocity profiles are found from
\begin{eqnarray}\label{crw:xi_dtPQ=u-c}
  \xi &\equiv&  \frac{x}{t} = u(\rho)-c_s(\rho), \\ \label{crw:u_dtPQ=}
  u(\rho) &=& u_D +c_{sJ}\left(1-\frac{v_D}{v_J}\right) + \int\limits_{\rho}^{\rho_J} c_s(\rho')\,d\ln\rho',
\end{eqnarray}
where $c_s(\rho)$ at $\rho\leq \rho_J$ must be calculated along the EQ expansion isentrope $s=s_J$; the outer edge velocity $u_e=u(0)$. In Fig.~\ref{f:3} the isentrope $s=s_J$ (dash-dotted black curve) visually merges with the corresponding segment $h_{PF}$ of the PF Hugoniot; note that the entropy increment $s_J-s_0$ in the PF shock (see Table~\ref{t:2}) is relatively small. By analogy with Eq.~(\ref{crw:Dxi_lvt=}), the effective width of the liquid-vapor tail is given by
\begin{equation}\label{crw:Dxi_lvtPF=}
  \Delta\xi_{lvt,PF} = c_{sJ}\Gamma_J,
\end{equation}
where $\Gamma_J$ is the fundamental gasdynamic derivative (\ref{crw:Gam=}) at point $J$. The values of $\Gamma_J$ and $\Delta\xi_{lvt,PF}$ for the illustrative cases 3 and 5 are listed in Table~\ref{t:2}. It should be emphasized that for all $s_0> s_{\ast}$ the PF-CRW profiles look qualitatively similar irrespective of how $s_0$ compares with $s_{cr}$, i.e.\ irrespective of whether the PF shock is of the vaporization or condensation type.

\subsection{Comparison between the EQ and PF solutions}

When comparing the EQ and PF profiles in CRW, we focus our attention on two main aspects that might be important for the interpretation of the experimental data. The feature that both solutions have in common is a horizontal shelf on spatial profiles, representing a layer of uniform flow. Only quantitative characteristics of the EQ and PF shelves are different. The main qualitative difference is that, except for the degenerate case of $s_0=s_{cr}$, the PF flow contains a discontinuity: it is either a density jump at the outer boundary with vacuum or a rarefaction shock downstream from the uniform layer.

When $s_0<s_{cr}$ and the expansion isentrope crosses the liquid binodal branch, the quantitative difference between the PF and EQ shelves is not large: it disappears in both limits of $p_B \to 0$ and $p_B \to 1$, and becomes maximum somewhere at $s_{\ast} \lesssim s_0 < s_{cr}$ (at $0.7 \lesssim p_B <1$ for our EOS). Generally, the PF shelf is broader, $\Delta\xi_{sh,PF} > \Delta\xi_{sh,EQ}$, but has a somewhat lower density ($\rho_E$ or $\rho_D$) than that ($\rho_B$) in the EQ case. Over the whole liquid branch, the values of both these parameters (cf.\ Tables \ref{t:1} and \ref{t:2}) differ by no more than a few percent. Also, the difference
\begin{equation}\label{crw:du_sh=}
  \delta u_{sh} = \int\limits^{\rho_B}_{\max\{\rho_E,\rho_D\}} c_s(\rho')\,d\ln\rho'
\end{equation}
between the flow velocities on the PF and EQ shelves, listed in the last row of Table~\ref{t:2}, remains small for $s_0< s_{cr}$. On the one hand, these  results confirm the statement by Inogamov \emph{et al.} \cite{Inogamov_Petrov.1999} that penetration into the metastable region of a superheated liquid should have only a minor effect on the unloading profiles. On the other hand, the same comparison for supercritical isentropes $s_0>s_{cr}$ (case~5 in Tables \ref{t:1} and \ref{t:2}) reveals that rapid condensation of the supersaturated vapor leads to a considerable distortion of the CRW profiles, which becomes more pronounced with the increasing entropy $s_0$.

The new quality, brought in by the PF approximation, is the presence of a discontinuity, which formally disappears only for the critical isentrope $s_0=s_{cr}$. Important practical implications of this discontinuity may be not so in the mechanical as in the optical properties of the rarefaction flows because in laser-driven experiments the absorption of the incoming laser light is sensitive to the density gradient in the expanding material. But even in this respect, the PF solution may still be practically indistinguishable from the EQ one if the effective width $\Delta\xi_{lvt,EQ}$ of the EQ-CRW tail, defined in Eq.~(\ref{crw:Dxi_lvt=}), is small enough --- which is definitely the case in the limit of $p_B \to 0$ along the liquid branch of the binodal.

Firstly, we note that, from the viewpoint of laser absorption, the PF density profiles for practically all subcritical isentropes $s_0<s_{cr}$ can be considered as ending with a sharp liquid surface of superheated liquid bordering on vacuum. Indeed, for $s_0< s_{\ast}$ ($p_B <0.74$, $\theta_B< 0.95$) this is literally the case, while for $s_0>s_{\ast}$, even as close to the critical point as $p_B = 0.86$, $\theta_B= 0.975$ (case~3 in Table~\ref{t:1}), we still observe a density jump by a large factor of $v_J/v_D =3.82$. The PF density jump disappears only in the immediate vicinity of the critical point, where the PF approximation becomes inaccurate anyway. Hence, the impact of metastability on the laser absorption can be claimed significant whenever the density gradient in the EQ liquid-vapor tail becomes of significance in this regard. This happens when the effective width of this tail, which in conventional units is given by
\begin{equation}\label{crw:h_lvt,EQ=}
  h_{lvt,EQ} =\Delta\xi_{lvt,EQ} \left(P_{cr}/\rho_{cr}\right)^{1/2} t,
\end{equation}
becomes comparable to the laser wavelength $\lambda$ [actually, $h_{lvt,EQ} \gtrsim (0.1$--0.2)${}\lambda$ would suffice]. Simple estimates indicate that for multi-nanosecond laser pulses this can often be the case for $\Delta\xi_{lvt,EQ} \gtrsim 0.03$, which corresponds to the boiling pressures $p_B \gtrsim 0.2$ within our EOS.

For supercritical isentropes $s_0>s_{cr}$ the difference between the PF and EQ density profiles is significant anyway, and an extra density jump, whose amplitude increases with $s_0$, only reinforces this conclusion.

\section{Conclusions}

The approximation of the phase-flip (PF) hydrodynamics, based on the assumption of instantaneous decay of the metastable liquid/vapor states at the spinodal, provides a possibility to explore the effects of metastability on the dynamic behavior of fluids with a liquid-gas phase transition without invoking more complex kinetic models. In this work, the first exact solution of the PF hydrodynamics is derived and investigated for the centered rarefaction wave (CRW) by unloading of a uniform layer into vacuum. Numerical results are obtained for a particular version of the generalized van der Waals EOS in the reduced form.

Of special interest for practical problems becomes a comparison of the PF solution with the simpler and more familiar EQ solution \cite{Sokol-Tint_Bialkowski.1998, Inogamov_Anisimov.1999, Zhao_Mentrelli.2011}, obtained for the same problem by using the fully equilibrium version of EOS in the two-phase region. One would expect that such a comparison should reveal the maximum possible influence of metastability in the problem considered. Here, having applied this approach to the classical problem of unloading into vacuum, we obtained the following results.

We find that the new qualitative feature, emerging in the PF solution, is the appearance of a flow discontinuity, i.e.\ of a rarefaction shock front. Nevertheless, so long as the boiling pressure (normalized by  $P_{cr}$) on a liquid expansion isentrope stays low $p_B \ll 1$, the difference between the continuous EQ and the discontinuous PF solutions remains, to all purposes, insignificant because a very steep drop in the EQ density profile downstream from the binodal shelf can hardly be distinguished from the step function in the PF profile. But as the boiling pressure approaches the critical value, metastability becomes an issue from the point of view of absorption and reflection of the laser light. In particular, for $s_0\leq s_{\ast}$ ($p_B\leq 0.74$ in our EOS) the PF solution predicts a sharp liquid surface at the boundary with vacuum, which for metals can be highly reflective according to the Fresnel formulae. The EQ solution, on the contrary, exhibits a smooth density slope along the disperse liquid-vapor tail, with the steepest gradient characterized by the length scale $h_{lvt,EQ}$ given by Eq.~(\ref{crw:h_lvt,EQ=}). With $\Delta\xi_{lvt,EQ} \approx 0.3$ for $s_0=s_{\ast}$ (case~2 in Table~\ref{t:1}) and $\left(P_{cr}/\rho_{cr}\right)^{1/2} \gtrsim 3\times 10^4$~cm/s for typical metals, we calculate $h_{lvt,EQ} \gtrsim 0.1$~$\mu$m for times $t \gtrsim 1$~ns, which would be comparable to the laser wavelength $\lambda$ in most experiments with nanosecond laser pulses. As a consequence, simulations of laser experiments, based on the EQ EOS, may significantly overestimate the laser absorption at low irradiation intensities near the ablation threshold. In particular, this could explain the discrepancy between the observed and predicted onset of the ablation regime in Ref.~\onlinecite{Kurilovich_Basko.2018}.

At the same time, there are many experiments with ultrashort laser pulses where the laser pulse terminates before the target begins to expand, and the key parameters for adequate interpretation of the experimental data become the width, mass and velocity of the plateau layer (the binodal/spinodal shelf) in the rarefaction wave\cite{Sokol-Tint_Bialkowski.1998, Basko_Krivokor.2017}. In this regard our analysis confirms and quantifies the earlier conclusion \cite{Inogamov_Petrov.1999} about the insignificant role of metastability, but under the condition that the expansion isentrope remains undercritical, $s_0< s_{cr}$. For supercritical isentropes  $s_0> s_{cr}$, where the rarefaction shock is a condensation jump, the PF approximation practically always predicts significant deviations from the EQ unloading profiles. This fact should be taken into account by interpretation of certain types of experiments where the EQ EOS is used.

\begin{acknowledgments}
The author is grateful to Anna Tauschwitz for many stimulating discussions. This work was  supported by the Russian Science Foundation through grant No.~14-11-00699-$\Pi$.
\end{acknowledgments}

\bibliography{Basko_pfhyd18-biblio}

\begin{thebibliography}{24}%
\makeatletter
\providecommand \@ifxundefined [1]{%
 \@ifx{#1\undefined}
}%
\providecommand \@ifnum [1]{%
 \ifnum #1\expandafter \@firstoftwo
 \else \expandafter \@secondoftwo
 \fi
}%
\providecommand \@ifx [1]{%
 \ifx #1\expandafter \@firstoftwo
 \else \expandafter \@secondoftwo
 \fi
}%
\providecommand \natexlab [1]{#1}%
\providecommand \enquote  [1]{``#1''}%
\providecommand \bibnamefont  [1]{#1}%
\providecommand \bibfnamefont [1]{#1}%
\providecommand \citenamefont [1]{#1}%
\providecommand \href@noop [0]{\@secondoftwo}%
\providecommand \href [0]{\begingroup \@sanitize@url \@href}%
\providecommand \@href[1]{\@@startlink{#1}\@@href}%
\providecommand \@@href[1]{\endgroup#1\@@endlink}%
\providecommand \@sanitize@url [0]{\catcode `\\12\catcode `\$12\catcode
  `\&12\catcode `\#12\catcode `\^12\catcode `\_12\catcode `\%12\relax}%
\providecommand \@@startlink[1]{}%
\providecommand \@@endlink[0]{}%
\providecommand \url  [0]{\begingroup\@sanitize@url \@url }%
\providecommand \@url [1]{\endgroup\@href {#1}{\urlprefix }}%
\providecommand \urlprefix  [0]{URL }%
\providecommand \Eprint [0]{\href }%
\providecommand \doibase [0]{http://dx.doi.org/}%
\providecommand \selectlanguage [0]{\@gobble}%
\providecommand \bibinfo  [0]{\@secondoftwo}%
\providecommand \bibfield  [0]{\@secondoftwo}%
\providecommand \translation [1]{[#1]}%
\providecommand \BibitemOpen [0]{}%
\providecommand \bibitemStop [0]{}%
\providecommand \bibitemNoStop [0]{.\EOS\space}%
\providecommand \EOS [0]{\spacefactor3000\relax}%
\providecommand \BibitemShut  [1]{\csname bibitem#1\endcsname}%
\let\auto@bib@innerbib\@empty
\bibitem [{\citenamefont {von~der Linde}, \citenamefont {Sokolowski-Tinten},\
  and\ \citenamefont {Bialkowski}(1997)}]{Linde_Sokow-Tinten.1997}%
  \BibitemOpen
  \bibfield  {author} {\bibinfo {author} {\bibfnamefont {D.}~\bibnamefont
  {von~der Linde}}, \bibinfo {author} {\bibfnamefont {K.}~\bibnamefont
  {Sokolowski-Tinten}}, \ and\ \bibinfo {author} {\bibfnamefont
  {J.}~\bibnamefont {Bialkowski}},\ }\bibfield  {title} {\enquote {\bibinfo
  {title} {{Laser-solid} interaction in the femtosecond time regime},}\ }\href
  {\doibase https://doi.org/10.1016/S0169-4332(96)00611-3} {\bibfield
  {journal} {\bibinfo  {journal} {Appl. Surf. Sci.}\ }\textbf {\bibinfo
  {volume} {109-110}},\ \bibinfo {pages} {1--10} (\bibinfo {year}
  {1997})}\BibitemShut {NoStop}%
\bibitem [{\citenamefont {Vinokhodov}\ \emph {et~al.}(2016)\citenamefont
  {Vinokhodov}, \citenamefont {Koshelev}, \citenamefont {Krivtsun},
  \citenamefont {Krivokorytov}, \citenamefont {Sidel'nikov}, \citenamefont
  {Medvedev}, \citenamefont {Kompanets}, \citenamefont {Mel'nikov},\ and\
  \citenamefont {Chekalin}}]{Vinokhodov_Koshelev.2016}%
  \BibitemOpen
  \bibfield  {author} {\bibinfo {author} {\bibfnamefont {A.~Y.}\ \bibnamefont
  {Vinokhodov}}, \bibinfo {author} {\bibfnamefont {K.~N.}\ \bibnamefont
  {Koshelev}}, \bibinfo {author} {\bibfnamefont {V.~N.}\ \bibnamefont
  {Krivtsun}}, \bibinfo {author} {\bibfnamefont {M.~S.}\ \bibnamefont
  {Krivokorytov}}, \bibinfo {author} {\bibfnamefont {Y.~V.}\ \bibnamefont
  {Sidel'nikov}}, \bibinfo {author} {\bibfnamefont {S.~V.}\ \bibnamefont
  {Medvedev}}, \bibinfo {author} {\bibfnamefont {V.~O.}\ \bibnamefont
  {Kompanets}}, \bibinfo {author} {\bibfnamefont {A.~A.}\ \bibnamefont
  {Mel'nikov}}, \ and\ \bibinfo {author} {\bibfnamefont {S.~V.}\ \bibnamefont
  {Chekalin}},\ }\bibfield  {title} {\enquote {\bibinfo {title} {Formation of a
  fine-dispersed liquid-metal target under the action of femto- and picosecond
  laser pulses for a laser-plasma radiation source in the extreme ultraviolet
  range},}\ }\href {\doibase 10.1070/QE2016v046n01ABEH015867} {\bibfield
  {journal} {\bibinfo  {journal} {Quantum Electronics}\ }\textbf {\bibinfo
  {volume} {46}},\ \bibinfo {pages} {23--28} (\bibinfo {year}
  {2016})}\BibitemShut {NoStop}%
\bibitem [{\citenamefont {Sokolowski-Tinten}\ \emph {et~al.}(1998)\citenamefont
  {Sokolowski-Tinten}, \citenamefont {Bialkowski}, \citenamefont {Cavalleri},
  \citenamefont {von~der Linde}, \citenamefont {Oparin}, \citenamefont
  {{Meyer-ter-Vehn}},\ and\ \citenamefont
  {Anisimov}}]{Sokol-Tint_Bialkowski.1998}%
  \BibitemOpen
  \bibfield  {author} {\bibinfo {author} {\bibfnamefont {K.}~\bibnamefont
  {Sokolowski-Tinten}}, \bibinfo {author} {\bibfnamefont {J.}~\bibnamefont
  {Bialkowski}}, \bibinfo {author} {\bibfnamefont {A.}~\bibnamefont
  {Cavalleri}}, \bibinfo {author} {\bibfnamefont {D.}~\bibnamefont {von~der
  Linde}}, \bibinfo {author} {\bibfnamefont {A.}~\bibnamefont {Oparin}},
  \bibinfo {author} {\bibfnamefont {J.}~\bibnamefont {{Meyer-ter-Vehn}}}, \
  and\ \bibinfo {author} {\bibfnamefont {S.~I.}\ \bibnamefont {Anisimov}},\
  }\bibfield  {title} {\enquote {\bibinfo {title} {Transient states of matter
  during short pulse laser ablation},}\ }\href {\doibase
  10.1103/PhysRevLett.81.224} {\bibfield  {journal} {\bibinfo  {journal} {Phys.
  Rev. Lett.}\ }\textbf {\bibinfo {volume} {81}},\ \bibinfo {pages} {224--227}
  (\bibinfo {year} {1998})}\BibitemShut {NoStop}%
\bibitem [{\citenamefont {Basko}\ \emph {et~al.}(2017)\citenamefont {Basko},
  \citenamefont {Krivokorytov}, \citenamefont {Vinokhodov}, \citenamefont
  {Sidelnikov}, \citenamefont {Krivtsun}, \citenamefont {Medvedev},
  \citenamefont {Kim}, \citenamefont {Kompanets}, \citenamefont {Lash},\ and\
  \citenamefont {Koshelev}}]{Basko_Krivokor.2017}%
  \BibitemOpen
  \bibfield  {author} {\bibinfo {author} {\bibfnamefont {M.~M.}\ \bibnamefont
  {Basko}}, \bibinfo {author} {\bibfnamefont {M.~S.}\ \bibnamefont
  {Krivokorytov}}, \bibinfo {author} {\bibfnamefont {A.~Y.}\ \bibnamefont
  {Vinokhodov}}, \bibinfo {author} {\bibfnamefont {Y.~V.}\ \bibnamefont
  {Sidelnikov}}, \bibinfo {author} {\bibfnamefont {V.~M.}\ \bibnamefont
  {Krivtsun}}, \bibinfo {author} {\bibfnamefont {V.~V.}\ \bibnamefont
  {Medvedev}}, \bibinfo {author} {\bibfnamefont {D.~A.}\ \bibnamefont {Kim}},
  \bibinfo {author} {\bibfnamefont {V.~O.}\ \bibnamefont {Kompanets}}, \bibinfo
  {author} {\bibfnamefont {A.~A.}\ \bibnamefont {Lash}}, \ and\ \bibinfo
  {author} {\bibfnamefont {K.~N.}\ \bibnamefont {Koshelev}},\ }\bibfield
  {title} {\enquote {\bibinfo {title} {Fragmentation dynamics of {liquid-metal}
  droplets under ultra-short laser pulses},}\ }\href
  {http://stacks.iop.org/1612-202X/14/i=3/a=036001} {\bibfield  {journal}
  {\bibinfo  {journal} {Laser Physics Letters}\ }\textbf {\bibinfo {volume}
  {14}},\ \bibinfo {pages} {036001} (\bibinfo {year} {2017})}\BibitemShut
  {NoStop}%
\bibitem [{\citenamefont {Inogamov}, \citenamefont {Anisimov},\ and\
  \citenamefont {Retfeld}(1999)}]{Inogamov_Anisimov.1999}%
  \BibitemOpen
  \bibfield  {author} {\bibinfo {author} {\bibfnamefont {N.~A.}\ \bibnamefont
  {Inogamov}}, \bibinfo {author} {\bibfnamefont {S.~I.}\ \bibnamefont
  {Anisimov}}, \ and\ \bibinfo {author} {\bibfnamefont {B.}~\bibnamefont
  {Retfeld}},\ }\bibfield  {title} {\enquote {\bibinfo {title} {Rarefaction
  wave and gravitational equilibrium in a two-phase liquid-vapor medium},}\
  }\href {\doibase 10.1134/1.558903} {\bibfield  {journal} {\bibinfo  {journal}
  {JETP}\ }\textbf {\bibinfo {volume} {88}},\ \bibinfo {pages} {1143--1150}
  (\bibinfo {year} {1999})}\BibitemShut {NoStop}%
\bibitem [{\citenamefont {Saurel}, \citenamefont {Petitpas},\ and\
  \citenamefont {Abgrall}(2008)}]{Saurel_Petitpas.2008}%
  \BibitemOpen
  \bibfield  {author} {\bibinfo {author} {\bibfnamefont {R.}~\bibnamefont
  {Saurel}}, \bibinfo {author} {\bibfnamefont {F.}~\bibnamefont {Petitpas}}, \
  and\ \bibinfo {author} {\bibfnamefont {R.}~\bibnamefont {Abgrall}},\
  }\bibfield  {title} {\enquote {\bibinfo {title} {Modelling phase transition
  in metastable liquids: application to cavitating and flashing flows},}\
  }\href {\doibase 10.1017/S0022112008002061} {\bibfield  {journal} {\bibinfo
  {journal} {J. Fluid Mech.}\ }\textbf {\bibinfo {volume} {607}},\ \bibinfo
  {pages} {313--350} (\bibinfo {year} {2008})}\BibitemShut {NoStop}%
\bibitem [{\citenamefont {Courant}\ and\ \citenamefont
  {Friedrichs}(1977)}]{Courant_Friedrichs1977}%
  \BibitemOpen
  \bibfield  {author} {\bibinfo {author} {\bibfnamefont {R.}~\bibnamefont
  {Courant}}\ and\ \bibinfo {author} {\bibfnamefont {K.}~\bibnamefont
  {Friedrichs}},\ }\href@noop {} {\emph {\bibinfo {title} {Supersonic Flow and
  Shock Waves}}},\ \bibinfo {series} {Applied mathematical sciences}\ No.\
  \bibinfo {number} {v. 21}\ (\bibinfo  {publisher} {Springer-Verlag},\
  \bibinfo {year} {1977})\BibitemShut {NoStop}%
\bibitem [{\citenamefont {Landau}\ and\ \citenamefont
  {Lifshitz}(1987)}]{LL-H87}%
  \BibitemOpen
  \bibfield  {author} {\bibinfo {author} {\bibfnamefont {L.}~\bibnamefont
  {Landau}}\ and\ \bibinfo {author} {\bibfnamefont {E.}~\bibnamefont
  {Lifshitz}},\ }\href@noop {} {\emph {\bibinfo {title} {Fluid Mechanics}}},\
  Course of theoretical physics\ (\bibinfo  {publisher} {Pergamon Press},\
  \bibinfo {year} {1987})\BibitemShut {NoStop}%
\bibitem [{\citenamefont {Zel'dovich}\ and\ \citenamefont
  {Raizer}(2012)}]{Zeldovich_Raizer2012}%
  \BibitemOpen
  \bibfield  {author} {\bibinfo {author} {\bibfnamefont {Y.~B.}\ \bibnamefont
  {Zel'dovich}}\ and\ \bibinfo {author} {\bibfnamefont {Y.~P.}\ \bibnamefont
  {Raizer}},\ }\href@noop {} {\emph {\bibinfo {title} {Physics of Shock Waves
  and High-Temperature Hydrodynamic Phenomena}}},\ Dover Books on Physics\
  (\bibinfo  {publisher} {Dover Publications},\ \bibinfo {year}
  {2012})\BibitemShut {NoStop}%
\bibitem [{\citenamefont {Martynyuk}(1991)}]{Martynyuk1991}%
  \BibitemOpen
  \bibfield  {author} {\bibinfo {author} {\bibfnamefont {M.~M.}\ \bibnamefont
  {Martynyuk}},\ }\bibfield  {title} {\enquote {\bibinfo {title} {Generalized
  van der {W}aals equation of state for liquids and gases},}\ }\href@noop {}
  {\bibfield  {journal} {\bibinfo  {journal} {Zh. Fiz. Khim.}\ }\textbf
  {\bibinfo {volume} {65}},\ \bibinfo {pages} {1716--1717} (\bibinfo {year}
  {1991})}\BibitemShut {NoStop}%
\bibitem [{\citenamefont {Martynyuk}(1993)}]{Martynyuk1993}%
  \BibitemOpen
  \bibfield  {author} {\bibinfo {author} {\bibfnamefont {M.~M.}\ \bibnamefont
  {Martynyuk}},\ }\bibfield  {title} {\enquote {\bibinfo {title} {Transition of
  liquid metals into vapor in the process of pulse heating by current},}\
  }\href {\doibase 10.1007/BF00566045} {\bibfield  {journal} {\bibinfo
  {journal} {International Journal of Thermophysics}\ }\textbf {\bibinfo
  {volume} {14}},\ \bibinfo {pages} {457--470} (\bibinfo {year}
  {1993})}\BibitemShut {NoStop}%
\bibitem [{\citenamefont {Basko}(2018)}]{Basko2018}%
  \BibitemOpen
  \bibfield  {author} {\bibinfo {author} {\bibfnamefont {M.~M.}\ \bibnamefont
  {Basko}},\ }\bibfield  {title} {\enquote {\bibinfo {title} {Generalized {van
  der Waals} equation of state for in-line use in hydrodynamic codes},}\ }\href
  {\doibase 10.20948/prepr-2018-112-e} {\bibfield  {journal} {\bibinfo
  {journal} {Keldysh Institute Preprints}\ }\textbf {\bibinfo {volume} {112}},\
  \bibinfo {pages} {28~p.} (\bibinfo {year} {2018})}\BibitemShut {NoStop}%
\bibitem [{\citenamefont {Grigoriev}\ and\ \citenamefont
  {Meilikhov}(1997)}]{Grigoriev1997}%
  \BibitemOpen
  \bibfield  {author} {\bibinfo {author} {\bibfnamefont {I.~S.}\ \bibnamefont
  {Grigoriev}}\ and\ \bibinfo {author} {\bibfnamefont {E.~Z.}\ \bibnamefont
  {Meilikhov}},\ }\href {https://books.google.ru/books?id=IYVUAAAAMAAJ} {\emph
  {\bibinfo {title} {Handbook of Physical Quantities}}}\ (\bibinfo  {publisher}
  {CRC-Press},\ \bibinfo {year} {1997})\BibitemShut {NoStop}%
\bibitem [{\citenamefont {Martynyuk}\ and\ \citenamefont
  {Balasubramanian}(1995)}]{Martynyuk_B1995}%
  \BibitemOpen
  \bibfield  {author} {\bibinfo {author} {\bibfnamefont {M.~M.}\ \bibnamefont
  {Martynyuk}}\ and\ \bibinfo {author} {\bibfnamefont {R.}~\bibnamefont
  {Balasubramanian}},\ }\bibfield  {title} {\enquote {\bibinfo {title}
  {Equation of state for fluid alkali metals: Binodal},}\ }\href {\doibase
  10.1007/BF01441919} {\bibfield  {journal} {\bibinfo  {journal} {International
  Journal of Thermophysics}\ }\textbf {\bibinfo {volume} {16}},\ \bibinfo
  {pages} {533--543} (\bibinfo {year} {1995})}\BibitemShut {NoStop}%
\bibitem [{\citenamefont {Landau}\ and\ \citenamefont
  {Lifshitz}(1996)}]{LL-SP96}%
  \BibitemOpen
  \bibfield  {author} {\bibinfo {author} {\bibfnamefont {L.~D.}\ \bibnamefont
  {Landau}}\ and\ \bibinfo {author} {\bibfnamefont {E.~M.}\ \bibnamefont
  {Lifshitz}},\ }\href@noop {} {\emph {\bibinfo {title} {{S}tatistical
  {P}hysics}}},\ \bibinfo {edition} {3rd}\ ed.\ (\bibinfo  {publisher}
  {Butterworth Heinemann},\ \bibinfo {year} {1996})\BibitemShut {NoStop}%
\bibitem [{\citenamefont {Skripov}(1974)}]{Skripov1974}%
  \BibitemOpen
  \bibfield  {author} {\bibinfo {author} {\bibfnamefont {V.~P.}\ \bibnamefont
  {Skripov}},\ }\href@noop {} {\emph {\bibinfo {title} {{M}etastable
  {L}iquids}}}\ (\bibinfo  {publisher} {Wiley, New York},\ \bibinfo {year}
  {1974})\BibitemShut {NoStop}%
\bibitem [{\citenamefont {Blander}\ and\ \citenamefont
  {Katz}(1975)}]{Blander_Katz1975}%
  \BibitemOpen
  \bibfield  {author} {\bibinfo {author} {\bibfnamefont {M.}~\bibnamefont
  {Blander}}\ and\ \bibinfo {author} {\bibfnamefont {J.~L.}\ \bibnamefont
  {Katz}},\ }\bibfield  {title} {\enquote {\bibinfo {title} {{B}ubble
  nucleation in liquids},}\ }\href {\doibase 10.1002/aic.690210502} {\bibfield
  {journal} {\bibinfo  {journal} {AIChE Journal}\ }\textbf {\bibinfo {volume}
  {21}},\ \bibinfo {pages} {833--848} (\bibinfo {year} {1975})}\BibitemShut
  {NoStop}%
\bibitem [{\citenamefont {Skripov}\ and\ \citenamefont
  {Skripov}(1979)}]{Skripov_Skripov1979}%
  \BibitemOpen
  \bibfield  {author} {\bibinfo {author} {\bibfnamefont {V.~P.}\ \bibnamefont
  {Skripov}}\ and\ \bibinfo {author} {\bibfnamefont {A.~V.}\ \bibnamefont
  {Skripov}},\ }\bibfield  {title} {\enquote {\bibinfo {title} {Spinodal
  decomposition (phase transitions via unstable states)},}\ }\href {\doibase
  10.1070/PU1979v022n06ABEH005571} {\bibfield  {journal} {\bibinfo  {journal}
  {Sov. Phys. Usp.}\ }\textbf {\bibinfo {volume} {22}},\ \bibinfo {pages} {389}
  (\bibinfo {year} {1979})}\BibitemShut {NoStop}%
\bibitem [{\citenamefont {Faik}\ \emph {et~al.}(2012)\citenamefont {Faik},
  \citenamefont {Basko}, \citenamefont {Tauschwitz}, \citenamefont
  {Iosilevskiy},\ and\ \citenamefont {Maruhn}}]{Faik_Basko.2012}%
  \BibitemOpen
  \bibfield  {author} {\bibinfo {author} {\bibfnamefont {S.}~\bibnamefont
  {Faik}}, \bibinfo {author} {\bibfnamefont {M.~M.}\ \bibnamefont {Basko}},
  \bibinfo {author} {\bibfnamefont {A.}~\bibnamefont {Tauschwitz}}, \bibinfo
  {author} {\bibfnamefont {I.}~\bibnamefont {Iosilevskiy}}, \ and\ \bibinfo
  {author} {\bibfnamefont {J.~A.}\ \bibnamefont {Maruhn}},\ }\bibfield  {title}
  {\enquote {\bibinfo {title} {Dynamics of volumetrically heated matter passing
  through the liquid-vapor metastable states},}\ }\href {\doibase
  10.1016/j.hedp.2012.08.003} {\bibfield  {journal} {\bibinfo  {journal} {High
  Energy Density Phys.}\ }\textbf {\bibinfo {volume} {8}},\ \bibinfo {pages}
  {349--359} (\bibinfo {year} {2012})}\BibitemShut {NoStop}%
\bibitem [{\citenamefont {Martynyuk}(1977)}]{Martynyuk1977}%
  \BibitemOpen
  \bibfield  {author} {\bibinfo {author} {\bibfnamefont {M.~M.}\ \bibnamefont
  {Martynyuk}},\ }\bibfield  {title} {\enquote {\bibinfo {title} {Phase
  explosion of a metastable fluid},}\ }\href {\doibase 10.1007/BF00754998}
  {\bibfield  {journal} {\bibinfo  {journal} {Combustion, Explosion and Shock
  Waves}\ }\textbf {\bibinfo {volume} {13}},\ \bibinfo {pages} {178--191}
  (\bibinfo {year} {1977})}\BibitemShut {NoStop}%
\bibitem [{\citenamefont {Thompson}(1971)}]{Thomson1971}%
  \BibitemOpen
  \bibfield  {author} {\bibinfo {author} {\bibfnamefont {P.~A.}\ \bibnamefont
  {Thompson}},\ }\bibfield  {title} {\enquote {\bibinfo {title} {A fundamental
  derivative in gasdynamics},}\ }\href {\doibase 10.1063/1.1693693} {\bibfield
  {journal} {\bibinfo  {journal} {Phys. Fluids (1958-1988)}\ }\textbf {\bibinfo
  {volume} {14}},\ \bibinfo {pages} {1843--1849} (\bibinfo {year}
  {1971})}\BibitemShut {NoStop}%
\bibitem [{\citenamefont {Inogamov}\ \emph {et~al.}(1999)\citenamefont
  {Inogamov}, \citenamefont {Petrov}, \citenamefont {Anisimov}, \citenamefont
  {Oparin}, \citenamefont {Shaposhnikov}, \citenamefont {von~der Linde},\ and\
  \citenamefont {ter Vehn}}]{Inogamov_Petrov.1999}%
  \BibitemOpen
  \bibfield  {author} {\bibinfo {author} {\bibfnamefont {N.~A.}\ \bibnamefont
  {Inogamov}}, \bibinfo {author} {\bibfnamefont {Y.~V.}\ \bibnamefont
  {Petrov}}, \bibinfo {author} {\bibfnamefont {S.~I.}\ \bibnamefont
  {Anisimov}}, \bibinfo {author} {\bibfnamefont {A.~M.}\ \bibnamefont
  {Oparin}}, \bibinfo {author} {\bibfnamefont {N.~V.}\ \bibnamefont
  {Shaposhnikov}}, \bibinfo {author} {\bibfnamefont {D.}~\bibnamefont {von~der
  Linde}}, \ and\ \bibinfo {author} {\bibfnamefont {J.~M.}\ \bibnamefont {ter
  Vehn}},\ }\bibfield  {title} {\enquote {\bibinfo {title} {Expansion of matter
  heated by an ultrashort laser pulse},}\ }\href {\doibase 10.1134/1.568029}
  {\bibfield  {journal} {\bibinfo  {journal} {JETP Lett.}\ }\textbf {\bibinfo
  {volume} {69}},\ \bibinfo {pages} {310--316} (\bibinfo {year}
  {1999})}\BibitemShut {NoStop}%
\bibitem [{\citenamefont {Zhao}\ \emph {et~al.}(2011)\citenamefont {Zhao},
  \citenamefont {Mentrelli}, \citenamefont {Ruggeri},\ and\ \citenamefont
  {Sugiyama}}]{Zhao_Mentrelli.2011}%
  \BibitemOpen
  \bibfield  {author} {\bibinfo {author} {\bibfnamefont {N.}~\bibnamefont
  {Zhao}}, \bibinfo {author} {\bibfnamefont {A.}~\bibnamefont {Mentrelli}},
  \bibinfo {author} {\bibfnamefont {T.}~\bibnamefont {Ruggeri}}, \ and\
  \bibinfo {author} {\bibfnamefont {M.}~\bibnamefont {Sugiyama}},\ }\bibfield
  {title} {\enquote {\bibinfo {title} {Admissible shock waves and shock-induced
  phase transitions in a van der {W}aals fluid},}\ }\href {\doibase
  10.1063/1.3622772} {\bibfield  {journal} {\bibinfo  {journal} {Phys. Fluids}\
  }\textbf {\bibinfo {volume} {23}},\ \bibinfo {pages} {086101} (\bibinfo
  {year} {2011})}\BibitemShut {NoStop}%
\bibitem [{\citenamefont {Kurilovich}\ \emph {et~al.}(2018)\citenamefont
  {Kurilovich}, \citenamefont {Basko}, \citenamefont {Kim}, \citenamefont
  {Torretti}, \citenamefont {Schupp}, \citenamefont {Visschers}, \citenamefont
  {Scheers}, \citenamefont {Hoekstra}, \citenamefont {Ubachs},\ and\
  \citenamefont {Versolato}}]{Kurilovich_Basko.2018}%
  \BibitemOpen
  \bibfield  {author} {\bibinfo {author} {\bibfnamefont {D.}~\bibnamefont
  {Kurilovich}}, \bibinfo {author} {\bibfnamefont {M.~M.}\ \bibnamefont
  {Basko}}, \bibinfo {author} {\bibfnamefont {D.~A.}\ \bibnamefont {Kim}},
  \bibinfo {author} {\bibfnamefont {F.}~\bibnamefont {Torretti}}, \bibinfo
  {author} {\bibfnamefont {R.}~\bibnamefont {Schupp}}, \bibinfo {author}
  {\bibfnamefont {J.~C.}\ \bibnamefont {Visschers}}, \bibinfo {author}
  {\bibfnamefont {J.}~\bibnamefont {Scheers}}, \bibinfo {author} {\bibfnamefont
  {R.}~\bibnamefont {Hoekstra}}, \bibinfo {author} {\bibfnamefont
  {W.}~\bibnamefont {Ubachs}}, \ and\ \bibinfo {author} {\bibfnamefont {O.~O.}\
  \bibnamefont {Versolato}},\ }\bibfield  {title} {\enquote {\bibinfo {title}
  {Power-law scaling of plasma pressure on laser-ablated tin microdroplets},}\
  }\href {\doibase 10.1063/1.5010899} {\bibfield  {journal} {\bibinfo
  {journal} {Phys. Plasmas}\ }\textbf {\bibinfo {volume} {25}},\ \bibinfo
  {pages} {012709} (\bibinfo {year} {2018})}\BibitemShut {NoStop}%
\end{thebibliography}%

\end{document}